\documentclass[aps,prc,groupedaddress,showpacs,manuscript]{revtex4}
\usepackage{graphicx}% Include figure files

\usepackage{colordvi}
\usepackage{color}

\begin{document}
%\begin{CJK*} {UTF8}{} %{GB} {gbsn}
%\preprint{APS/123-QED}

\title{The density- and isospin- dependent $\Delta$-formation cross section and its decay width}
%\thanks{A footnote to the article title}%

\author {Qingfeng Li$\, ^{1,2}$\footnote{E-mail address: liqf@zjhu.edu.cn} and
Zhuxia Li$\, ^{3}$}

\affiliation{
1) School of Science, Huzhou University, Huzhou 313000, P.R. China \\
2) Institute of Modern Physics, Chinese Academy of Sciences,  Lanzhou 730000,  P. R. China\\
3) China Institute of Atomic Energy, Beijing 102413, P. R. China\\
\\
 }
\date{\today}

\begin{abstract}
The energy-, density-, and isospin-dependent $\Delta$-formation cross section $\sigma_{N\pi \rightarrow \Delta}^*$ and $\Delta$-decay width are calculated based on the relativistic BUU approach in which the effective mass splitting of nucleon and $\Delta$ baryons in isospin-asymmetric matter is considered by the inclusion of the $\delta$ meson exchange in the effective Lagrangian density. With the density-dependent couplings for baryons of F. Hofmann et al., the $\sigma_{N\pi \rightarrow \Delta}^*$ is decreased (increased) moderately with increasing density with (without) the consideration of the density dependent pion effective mass. Meanwhile, if the invariant mass of the system is not far from the $\Delta$ pole mass, the $\Delta$-decay width is also weakly dependent on density. The mass splitting effect of differently charged nucleon and $\Delta$ baryons on $\sigma_{N\pi \rightarrow \Delta}^*$ are found to be more obvious than that of pion mesons but much weaker than the mass splitting in the hard $\Delta$ production channel $NN\rightarrow N\Delta$. Further, the largest mass-splitting influence is seen in the $\pi^-p\rightarrow \Delta^0$ and $\pi^+n\rightarrow \Delta^+$ channels but not in the production of $\Delta^-$ and $\Delta^{++}$ isobars.

\end{abstract}

% insert suggested PACS numbers in braces on next line

% insert suggested keywords - APS authors don't need to do this

%\maketitle must follow title, authors, abstract, \pacs, and \keywords
\maketitle
%\end{CJK*}

The existence of dubbed isospin so as to the current hot topic on the search for the stiffness of the symmetry energy at whatever nuclear densities visibly provides us a nice and uncommon opportunity for a deeper and broader understanding of the finite nuclei, the nuclear matter as well as even the evolution of the whole universe. Artificially, a situation far-away from the so-called normal nuclear density ($\rho_0$) can be achieved through a real nucleus-nucleus collision with the help of an accelerator and/or a simulated numerical calculations run by a transport model. But, the valuable information of the symmetry energy can only be extracted by a careful comparison between both sides mentioned above. So far, the experimental conditions have been greatly improved (e.g., the more neutron-rich projectile-target combination, the more rigid windows for more sensitive probes with higher precision) so that more and more new experiments related to the determination of symmetry energy have been performed or are coming soon.  Meanwhile, more and more well polished microscopic transport models profitted from the progresses of related fundamentally theoretical issues are available and strongly supporting us for a more reliable information of the symmetry energy at not only subnormal but also, more urgently, supranormal densities. Although, it is known that the large uncertainty exhibited in the results of several different model calculations has not been thoroughly clarified \cite{Xiao:2008vm,Feng:2009am,Trautmann:2009kq,Russotto:2011hq} which, meanwhile, are being more systematically checked by the Transport Simulation Code Evaluation Project \cite{Xu:2016lue,Zhang:2017esm}.

It is easy to understand that the yield ratio of charged $\pi^-$ and $\pi^+$ mesons is often selected to be a sensitive probe for extracting the stiffness of symmetry energy at supranormal densities since the pions are mainly produced from high densities where their parent particle $\Delta$s might be produced afore and decay soon later. Moreover, the bombarding energy of the projectile for the production of pions is the lowest when comparing with those for other new-produced stable particles so that the nuclear situation consisting of pure nucleons (neutrons and protons) has not been destroyed seriously. This is favorable since uncertainties from other newly produced particles can be largely avoided, except for its parent particle, $\Delta$. At SIS energies, the $\Delta(1232)$ resonance is the main source of the pion production. And, its evolution in nuclear medium, as well as for the accompanying nucleon and pion, should be investigated more objectively and systematically. There are two channels, namely $NN\rightarrow N\Delta$ ($\emph{hard}-\Delta$ production) and $N\pi \rightarrow \Delta$ ($\emph{soft}-\Delta$ production), which are crucial for the production and reabsorption of pions. And, similar to treatments on nucleons, the density- and isospin-dependent medium modifications should be accordingly considered for $\Delta$s and $\pi$s in the isospin-asymmetric nuclear medium, which, hence, are  reflected mainly in three aspects: the mean field, the production threshold, and the cross section. In recent years these aspects have been investigated for the $NN\rightarrow N\Delta$ process in medium  \cite{Prassa:2007zw,Song:2015hua,Li:2016xix} and it is found that all of them are important for the production of $\Delta$s. Especially, in Ref.~\cite{Li:2016xix} we found that the medium effect of $N\Delta$ inelastic cross sections in isospin-asymmetric nuclear medium are even different from each other for all individual channels if the effective mass splitting of the nucleons and $\Delta$s is taken into account within the framework of the relativistic Boltzmann-Uehling-Uhlenbeck (RBUU) approach.

After the hard $\Delta$ production process via $NN\rightarrow N\Delta$, besides its detailed balance process, the $\Delta$s will decay through $\Delta\rightarrow N\pi$ and the following $\Delta\rightarrow N\pi \rightarrow \Delta$ loop sustains for a relatively long time before the final freeze-out of pions (see, e.g., Ref.~\cite{Bass:1998ca} and the references therein). Therefore, the medium modifications on the $\Delta$-formation cross section $\sigma_{N\pi \rightarrow \Delta}^*$ and $\Delta$-decay process including its width $\Gamma^*$ should be taken further account. In Ref.~\cite{Zhang:2017mps}, including both the isospin-dependent pion s-wave (based on the chiral perturbation theory) and p-wave potentials (based on the $\Delta$-hole model) and the in-medium effects on both $\Delta$ production threshold and cross sections  $\sigma_{N\pi \rightarrow \Delta}^*$ and  $\sigma_{NN \rightarrow N\Delta}^*$ in the extended relativistic Vlasov-Uehling-Uhlenbeck (RVUU), a soft nuclear symmetry energy with a slope parameter $L=59$ MeV is also predicted from the comparison to the experimental $\pi^-/\pi^+$ ratio in Au+Au collisions at the beam energy $E_{lab}=400$ MeV$/$nucleon. Especially, it was claimed that the mass-splitting effect of pions from the s-wave potential suppresses to some extent the $\pi^-/\pi^+$ ratio. Hence, it is interesting to see how much the mass-splitting effect of other two participants, the nucleon and the $\Delta$ resonance, influences the  $N\pi\rightleftharpoons \Delta$ process, which will be focused in the current work.

This issue can be examined immediately based on three previous works by our group \cite{Mao:1998pr,Li:2016xix} and Ko's group \cite{Zhang:2017mps}. More specifically, the formalisms for the $N\pi \rightarrow \Delta$ cross section and the $\Delta$-decay width as well as the density-dependent pion dispersion relation (so as to the density- and momentum-dependent but isospin-independent pion effective mass) are taken directly from Ref.~\cite{Mao:1998pr} where those for medium (both density- and isospin-dependent) modifications on effective masses of neutrons, protons and $\Delta$ isobars are from Ref.~\cite{Li:2016xix}. We notice that the $\Delta$-decay width was also studied by J. Gegelia et al. in Ref.~\cite{Gegelia:2016pjm} recently based on the baryon chiral perturbation theory. For the consistence, in this work we adopt the same formalism as in Ref.~\cite{Mao:1998pr}. Note that a mass-dependent coupling strength for the $N-\Delta-\pi$ vertex was introduced in order for a better description of the cross section at free space and is also in use for the current work. As for the splitting effect on pions, the same definition from the s-wave pion self-energy expressed in Ref.~\cite{Kaiser:2001bx} and used in Ref.~\cite{Zhang:2017mps} is adopted for simplicity, which, however, causes somewhat inconsistency in form between the two theoretical approaches since the pion mass splitting is not derived directly from the current RBUU approach. This problem is kept in this work since, as stated above, we concentrate on the mass-splitting effects of nucleon and $\Delta$ baryons.  It is seen that in the neutron-rich matter the trend of pion effective masses is $m_{\pi^-}^*>m_{\pi^0}^*>m_{\pi^+}^*$ and contrary to that of nucleon masses ($m_p^*>m_n^*$) as well as $\Delta$ masses ($m_{\Delta^{++}}^*>m_{\Delta^+}^*>m_{\Delta^0}^*>m_{\Delta^-}^*$). Further, factors for six individual channels of the $N\pi \rightarrow \Delta$ cross section, which are from corresponding Clebsch-Gordan coefficients of the isospin coupling of a pion with nucleon and $\Delta$ resonance, are also considered.

\begin{figure}[htbp]
\centering
\includegraphics[angle=0,width=0.9\textwidth]{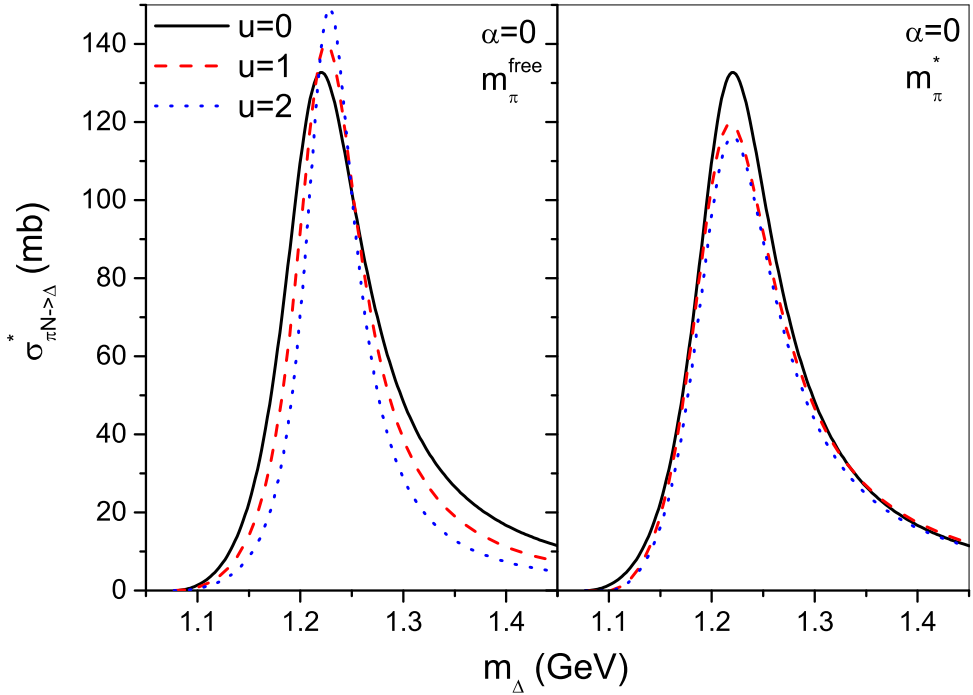}
\caption{\label{fig1} (Color online) Density-dependent (when isospin asymmetry $\alpha=0$) $\Delta$-formation cross sections with free pion mass (left) and with effective pion mass (right). }
\end{figure}

\begin{figure}[htbp]
\centering
\includegraphics[angle=0,width=0.9\textwidth]{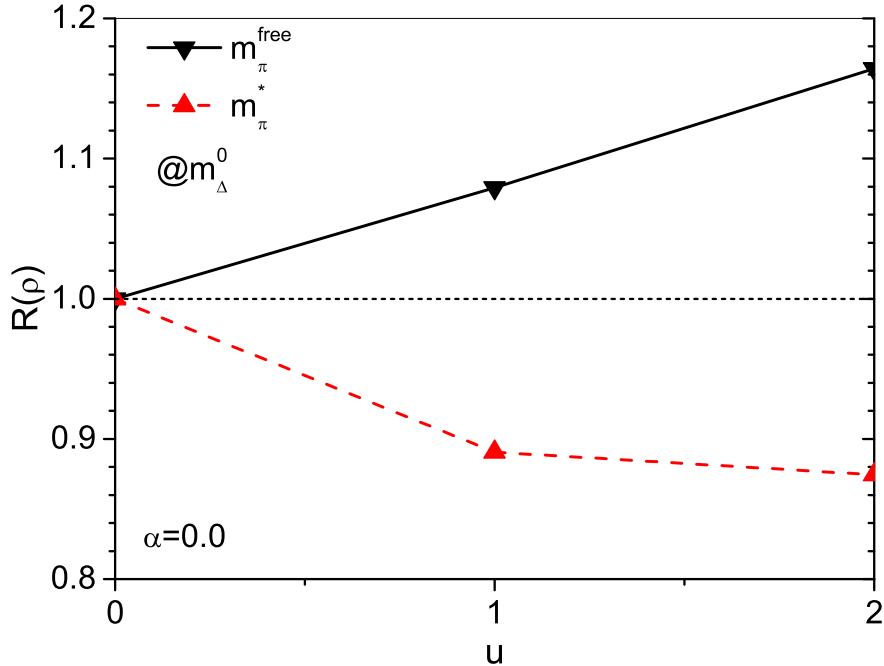}
\caption{\label{fig2} (Color online) The ratio $R(\rho)=\sigma^*/\sigma^{free}$ at the $\Delta$ pole mass as a function of the reduced density. The horizontal dotted line represents unity.}
\end{figure}

\begin{figure}[htbp]
\centering
\includegraphics[angle=0,width=0.9\textwidth]{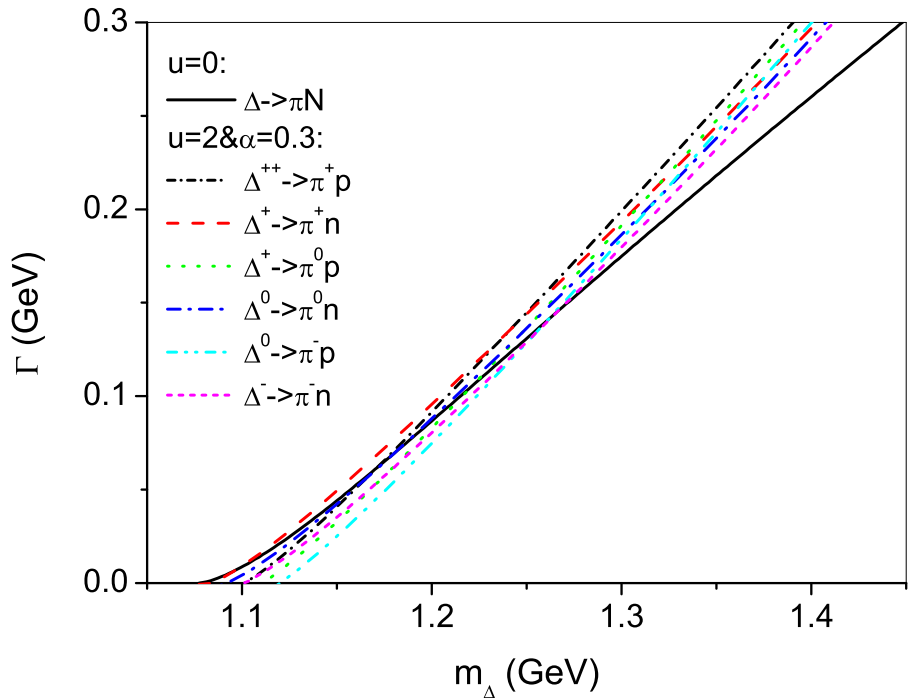}
\caption{\label{fig3} (Color online) The $\Delta$-decay widths at two densities u=0 and 2. At $u=2$ and $\alpha=0.3$ the splitted widths are shown by bunched lines. }
\end{figure}

Fistly, we show in Fig.~\ref{fig1} the in-medium $\Delta$-formation cross section for the isospin-symmetric matter ($\alpha=(\rho_n-\rho_p)/\rho=0$). Three reduced baryon densities $u$($=\rho/\rho_0$), 0, 1 and 2, are selected for calculations without (left plot) and with (right) the consideration of the medium modification on pions. Here, a set of parametrization for nucleon and $\Delta$ baryons coupling to $\sigma$, $\omega$, $\delta$ and $\rho$ mesons taken from Ref.~\cite{Hofmann:2000vz} is adopted, which, different from that used in Ref.~\cite{Mao:1998pr}, is density dependent, while for the couplings to $\pi$ meson, $g_{NN}^\pi=f_{\pi}/m_\pi$ and $g_{N\Delta}^\pi=f_{\pi}^*/m_\pi$, as well the parameters $f_{\pi}^{2}/4\pi=0.08$ and $f_{\pi}^{*2}/4\pi=0.362$ are taken from Ref.~\cite{Mao:1998pr}. A quite similar density dependence of the $\Delta$-formation cross section for both cases compared to those obtained in Ref.~\cite{Mao:1998pr} is reappeared. I.e., if the medium modification on pions is switched off and with the increase of density, the cross section is somewhat enhanced near the $\Delta$ pole mass ($m_\Delta^0=1.232$ GeV) and suppressed at other region. But, if the medium modification of pions is on, the cross section is contrarily decreased near the $m_\Delta^0$. This phenomenon can be understood mainly for the reason that both parametrization sets lead to the quite similar reduced effective nucleon mass $m^*/m_0$ at normal density; it is 0.55 in the current work while it is 0.6 in Ref.~\cite{Mao:1998pr}.

Fig.~\ref{fig2} further depicts the density-dependent ratio $R(\rho)=\sigma^*/\sigma^{free}$ at $m_\Delta^0$ as a function of the density. It is seen clearly that, at even two times of the normal density, the enhancement or suppression is not more than 16$\%$ of the free one. This weak density dependence of the $N\pi \rightarrow \Delta$ cross section is obviously different from that of the $NN\rightarrow N\Delta$ cross section calculated in the previous work ~\cite{Li:2016xix}, where the suppression factor is much smaller. Furthermore, if the invariant mass of the system is not far from the $m_\Delta^0$, the density dependence of the $\Delta$-decay width is also relatively weak, which is demonstrated in Fig.~\ref{fig3}. It should be noticed that the weak density dependence of both the $\Delta$-formation cross section shown in Fig.~\ref{fig1} and the $\Delta$-decay width shown in Fig.~\ref{fig3} is to some extent different from those calculated in Ref.~\cite{Zhang:2017mps} where a larger density dependence is seen. This difference might be due to the consideration of so-called short-range correlations in the nonrelativistic model via the Migdal parameter $g\prime$  ~\cite{Mao:1998pr,Brown:1975di}, which, however, is still an open problem for the  relativistic calculation.

\begin{figure}[htbp]
\centering
\includegraphics[angle=0,width=0.9\textwidth]{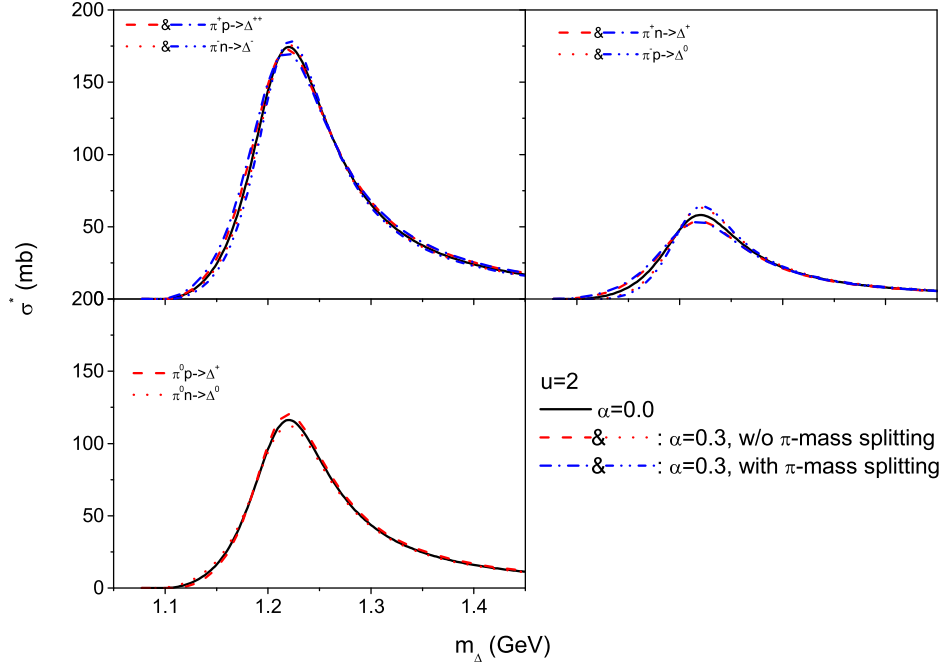}
\caption{\label{fig4} (Color online) The individual in-medium cross sections of the $N\pi\rightarrow \Delta$ process for $u=2$ and for two $\alpha$ values 0 and 0.3. When $\alpha=0.3$, two cases without and with the consideration of the $\pi$-mass splitting are compared to each other.}
\end{figure}

It is also seen from Fig.~\ref{fig3} that the mass-splitting effect on the $\Delta$-decay width is weak. E.g., at the $m_\Delta^0$ where the $\Gamma_0$ at free space is 115 MeV, the smallest width from the $\pi^-p\rightarrow \Delta^0$ is about 109 MeV while the largest one from the $\pi^+n\rightarrow \Delta^+$ is about 126 MeV at $u=2$ and $\alpha=0.3$. Here, the decay width is obtained from an initial-state averaged calculation of the $N\pi\rightarrow \Delta$ process so that the effect of the Clebsch-Gordan coefficient disappears and variances in width come purely from the mass-splitting effects of nucleon and $\Delta$ baryons (the pion-mass splitting is not considered so far). Therefore, it is worth to further check the mass-splitting effects of both nucleon and $\Delta$ baryons and pion mesons on the individual cross sections of all isospin-separated channels. These six cross sections are calculated at $u=2$ and are divided into three groups with the same Clebsch-Gordan coefficients, which are shown in each plot of Fig.~\ref{fig4} separately. Therefore, the factors in results of the upper-left, upper-right, and bottom-left plots are exactly 3/2, 1/2 and 1 respectively, which are hinted by the solid lines when $\alpha=0$. With  $\alpha\neq 0$, this relation is lost due to the effect of the mass splittings. If $\alpha=0.3$ and the baryon-mass splitting is considered but the $\pi$-mass splitting is not considered, the largest splitting effect is seen on the $\pi^+n\rightarrow \Delta^+$ and $\pi^-p\rightarrow \Delta^0$ cross sections. This is since the largest difference in mass modifications happens in these two channels when comparing the ingoing nucleon to the outgoing $\Delta$ resonance, while the transition probability of the $N\pi \rightarrow \Delta$ process is proportional to the term $(m_N^*-m_\Delta^*)^2-m_\pi^{*2}$ as obtained in Ref.~\cite{Mao:1998pr}.  When the $\pi$-mass splitting is further taken into account, the cross sections are not visibly varied. It had been seen in Ref.~\cite{Zhang:2017mps} that the difference between $\pi^+p\rightarrow\Delta^{++}$ and $\pi^-n\rightarrow\Delta^{-}$ cross sections at $u=2$ and $\alpha=0.2$ with both s- and p-wave potentials, where only the $\pi$-mass splitting is considered, is almost invisible as well. This is clear since the free mass of pion meson is much smaller than those of nucleon and $\Delta$ baryons, meanwhile, the effective mass difference of pion isobars via the self-energy is small. Hence, the decrease of the $\pi^-/\pi^+$ ratio seen in Ref.~\cite{Zhang:2017mps} originates mainly from the splitted potentials of pions but not from its follow-up splitted cross sections.

\begin{figure}[htbp]
\centering
\includegraphics[angle=0,width=0.9\textwidth]{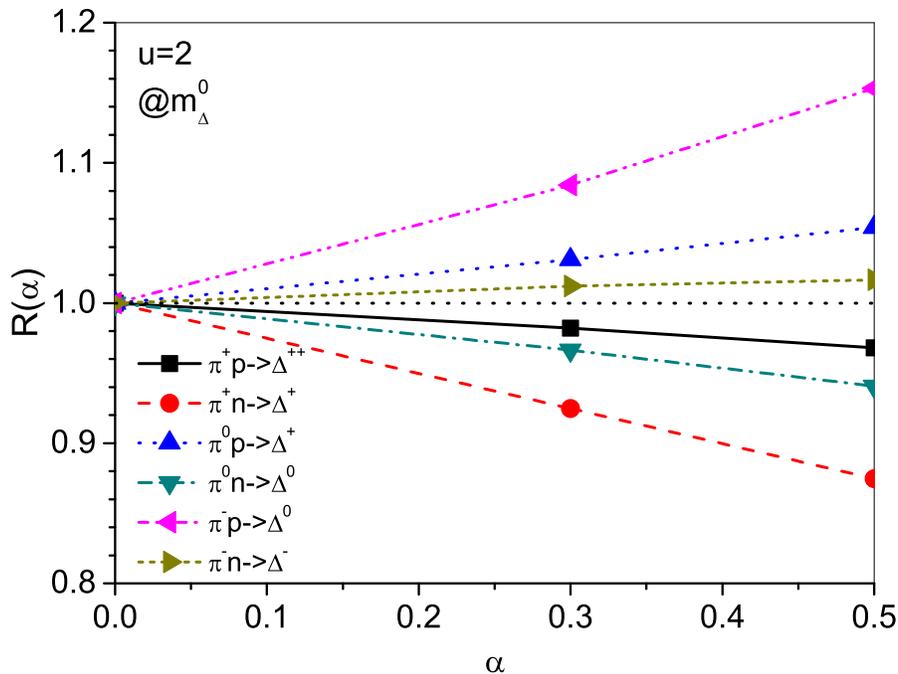}
\caption{\label{fig5} (Color online) The $R(\alpha)=\sigma^*(\alpha)/\sigma^*(\alpha=0)$ ratios of all channels at $m_\Delta^0$ as a function of the isospin asymmetry for $u=2$. The horizontal dotted line represents unity. }
\end{figure}

Similar to Fig.~\ref{fig2}, we show in Fig.~\ref{fig5} another isospin-dependent ratio $R(\alpha)=\sigma^*(\alpha)/\sigma^*(\alpha=0)$ of all channels at $m_\Delta^0$ (here the pion-mass splitting is neglected). First, ratios for three channels are enhanced while other three ratios are suppressed monotonously with increasing $\alpha$, and the sequence is regular as well: $R(\alpha, \pi^-p\rightarrow \Delta^0)>R(\alpha,\pi^0p\rightarrow \Delta^+)>R(\alpha,\pi^-n\rightarrow \Delta^-)>1>R(\alpha,\pi^+p\rightarrow \Delta^{++})>R(\alpha,\pi^0n\rightarrow \Delta^0)>R(\alpha, \pi^+n\rightarrow \Delta^+)$. At $u=2$, $R(\alpha)$=1.15, 1.05, 1.02, 0.97, 0.94, and 0.87, respectively. If we compare these ratios from the $N\pi\rightarrow \Delta$ process to those from the $NN\rightarrow N\Delta$ process in Ref.~\cite{Li:2016xix} for the production of same $\Delta$ isobar, it is interesting to see that, as a whole, the mass-splitting effect in $N\pi\rightarrow \Delta$ process is much weaker than that in $NN\rightarrow N\Delta$ process. And, the influence on $\Delta^-$ and $\Delta^{++}$ production channels from $N\pi\rightarrow \Delta$ ($NN\rightarrow N\Delta$) process is the weakest (strongest) and the order for both channels is also on the contrary.

To summarize, in this work, the energy-, density-, and isospin-dependent $\Delta$-formation cross section and $\Delta$-decay width are calculated based on the RBUU approach in which the effective mass splitting of nucleon and $\Delta$ baryons in isospin-asymmetric matter is considered.  The mass splitting effect of $\pi$ meson is taken from its s-wave self energy based on the chiral perturbation theory. It is found that the density dependence on $\sigma_{\pi N\rightarrow\Delta}^*$ is moderate, which is enhanced (with the consideration of the free $\pi$ mass) or suppressed (with the consideration of the density-dependent $\pi$ mass) by less than 20$\%$ at the $\Delta$ pole mass and at even two times of the normal density. Meanwhile, if the invariant mass of the system is not far from the $\Delta$ pole mass, the $\Delta$-decay width is also weakly dependent on density. As for the mass-splitting effect on $\sigma_{\pi N\rightarrow\Delta}^*$ which is mainly from the mass-splittings of nucleon and $\Delta$ baryons, the enhancement or suppression of individual cross sections is also moderate and less than 20$\%$ at the $\Delta$ pole mass and at two times of the normal density. And, the largest mass-splitting influence is reflected in the $\pi^-p\rightarrow \Delta^0$ and $\pi^+n\rightarrow \Delta^+$ processes but not in the production of $\Delta^-$ and $\Delta^{++}$ isobars.

Therefore, for the two main $\Delta$ production channels, namely $NN\rightarrow N\Delta$ and $N\pi\rightarrow \Delta$, at SIS energies, it is found that both the density- and the isospin- dependent cross sections of the $N\pi\rightarrow \Delta$ process are much weaker than those of the $NN\rightarrow N\Delta$ process. In the next step for numerical simulations  based on a microscopic transport model such as the ultrarelativistic quantum molecular dynamics (UrQMD) ~\cite{Bass:1998ca,Li:2011zzp,Wang:2013wca}, the medium modifications on  the $NN\rightarrow N\Delta$ process ought to be received more attention. And, it should be noticed that the currently splitting effect of individual channels in $\sigma_{\pi N\rightarrow\Delta}^*$ is enlarged due to the new inclusion of both nucleon- and $\Delta$- mass splittings if we compare it to the one with only the $\pi$-mass splitting. Its subsequent effect on the production of pions as well as the $\pi^-/\pi^+$ yield ratios needs to be examined as well. In addition, the effect of the short-range correlations on cross sections deserves further investigation.

\begin{acknowledgments}
We wish to thank H. St$\ddot{o}$cker, M. Bleicher, S. Schramm, and W. Trautmann for valuable discussions. The work is financially supported in part by the National
Natural Science Foundation of China (Nos. 11375062, 11647306, and 11475262). Q.L. acknowledges the warm hospitality of
FIAS institute during the stay in the middle of the year 2017.
\end{acknowledgments}

\end{document}